\documentclass[conference]{IEEEtran}
\IEEEoverridecommandlockouts
\usepackage{array,multirow,makecell}
\usepackage[T1]{fontenc}
\usepackage{multirow}
\usepackage{cite}
\usepackage{amsmath,amssymb,amsfonts}
\usepackage{algpseudocode}
\usepackage[utf8x]{inputenc}
\usepackage{tcolorbox}
\tcbuselibrary{listings}
\usepackage{ragged2e}
\usepackage{tcolorbox}
\usepackage{algorithm}
\usepackage{colortbl}
\usepackage{pgfplots} 
\usepgfplotslibrary{groupplots}
\pgfplotsset{compat=1.12}
\usepackage{pifont}% http://ctan.org/pkg/pifont
\newcommand{\cmark}{\ding{51}}%

\usetikzlibrary{arrows,shapes,positioning,shadows,trees}
\usepackage{rotating, adjustbox}
\newfloat{steps}{htbp}{lop}
\floatname{steps}{STEPS}
\IEEEoverridecommandlockouts
\usepackage{graphicx}
\usepackage{textcomp}
\usepackage{tcolorbox}
\usepackage[hyphens]{url}
\usepackage{amsthm}
\usepackage{cite}
\usepackage{pifont}
\usepackage{xspace}

    \makeatletter % changes the catcode of @ to 11
\newcommand{\linebreakand}{%
  \end{@IEEEauthorhalign}
  \hfill\mbox{}\par
  \mbox{}\hfill\begin{@IEEEauthorhalign}
}

\title{Revolutionizing Cyber Threat Detection with Large
Language Models: A privacy-preserving BERT-based Lightweight
Model for IoT/IIoT Devices}
\author{
\IEEEauthorblockN{Mohamed~Amine~Ferrag, Mthandazo Ndhlovu, Norbert Tihanyi, Lucas C. Cordeiro, \\ Merouane Debbah, Thierry Lestable and Narinderjit Singh Thandi}
 \IEEEauthorblockA{Technology Innovation Institute, 9639 Masdar City, Abu Dhabi, UAE}
 \IEEEauthorblockA{Email: firstname.lastname@tii.ae}
 }

\begin{document}
\maketitle

\begin{abstract}
The field of Natural Language Processing (NLP) is currently undergoing a revolutionary transformation driven by the power of pre-trained Large Language Models (LLMs) based on groundbreaking Transformer architectures. As the frequency and diversity of cybersecurity attacks continue to rise, the importance of incident detection has significantly increased. IoT devices are expanding rapidly, resulting in a growing need for efficient techniques to autonomously identify network-based attacks in IoT networks with both high precision and minimal computational requirements. This paper presents SecurityBERT, a novel architecture that leverages the Bidirectional Encoder Representations from Transformers (BERT) model for cyber threat detection in IoT networks. During the training of SecurityBERT, we incorporated a novel privacy-preserving encoding technique called  Privacy-Preserving Fixed-Length Encoding (PPFLE). We effectively represented network traffic data in a structured format by combining PPFLE with the Byte-level Byte-Pair Encoder (BBPE) Tokenizer. Our research demonstrates that SecurityBERT outperforms traditional Machine Learning (ML) and Deep Learning (DL) methods, such as Convolutional Neural Networks (CNNs) or Recurrent Neural Networks (RNNs), in cyber threat detection. Employing the Edge-IIoTset cybersecurity dataset, our experimental analysis shows that SecurityBERT achieved an impressive 98.2\% overall accuracy in identifying fourteen distinct attack types, surpassing previous records set by hybrid solutions such as GAN-Transformer-based architectures and CNN-LSTM models. With an inference time of less than 0.15 seconds on an average CPU and a compact model size of just 16.7MB, SecurityBERT is ideally suited for real-life traffic analysis and a suitable choice for deployment on resource-constrained IoT devices\footnote{This paper has been accepted for publication in IEEE Access: http://dx.doi.org/10.1109/ACCESS.2024.3363469}.
\end{abstract}

\section{Introduction}
\label{sec:introduction}
%---------------------------------------------
According to a Statista report~\cite{Statista}, it is projected that the global number of Internet of Things (IoT) connected devices could potentially reach 30 billion by the year 2030. With the rise in the number of IoT devices, there is also a growing incidence of cyber threats, posing substantial challenges to the security of diverse systems and networks~\cite{moustafa2023explainable}. As adversaries consistently evolve their tactics, the need for advanced and effective detection mechanisms becomes paramount. Manual detection methods and conventional approaches are becoming outdated, and various Machine Learning (ML) techniques have emerged, combating these new threats more effectively. In this context, Natural Language Processing (NLP) techniques are gaining attention as a promising approach for cyber threat detection~\cite{silvestri2023machine}. Among these techniques, the Bidirectional Encoder Representations from Transformers (BERT) model~\cite{thapa2022transformer}, a pre-trained transformer-based language model, has achieved remarkable success in several NLP applications. By exploiting BERT's contextual understanding, security researchers have found unique techniques to handle diverse cybersecurity concerns~\cite{devlin2018bert}. Researchers have recently been exploring using BERT and pre-trained language models in a wide range of cybersecurity applications, including malware detection in Android applications, identification of spam emails, intrusion detection in automotive systems, and anomaly detection in system logs~\cite{Rahali2021MalBERTMD, seyyar2022attack, Chen2022BERTLogAD}.  Network-based traffic, such as port scans and packet floods, primarily consists of numerical data rather than textual information. This characteristic poses a challenge when attempting to leverage models like BERT to understand the semantic relationships between various types of network packets. While employing complex Large Language Models (LLMs) with billions of parameters can improve threat detection accuracy, their extensive computational needs render them impractical for implementation in embedded devices.

We present \texttt{SecurityBERT}, a novel lightweight privacy-preserving architecture for cyber threat detection in IoT networks. By employing a dedicated encoding technique designed for this specific purpose, we surpassed the performance of all existing ML algorithms and models in cyber threat detection. During the design of \texttt{SecurityBERT}, we had three primary goals in mind:
\begin{itemize}
    \item  To create an exceptionally compact model capable of executing rapid inferences without causing noticeable delays. This design choice enables real-time traffic analysis and facilitates embedding the model in IoT devices;
    \item To maintain the confidentiality of the extracted network data, ensuring that classification can be performed on untrusted servers;
     \item To surpass the accuracy levels of previous ML models in this field.
\end{itemize}
Achieving superior accuracy compared to existing hybrid solutions has proven a significant challenge in our architectural design. Striking the right balance is crucial. If the architecture becomes overly complex, it may become impractical for real-life traffic analysis. Conversely, if the model is overly simplistic, it may not provide the necessary accuracy for effective multi-classification, thus hindering its overall performance. Our original contributions are as follows:
\begin{itemize}    
    \item Our research introduces a novel privacy-preserving encoding approach called Privacy-Preserving Fixed-Length Encoding (PPFLE). By combining PPFLE with the Byte-level Byte-Pair Encoder (BBPE) tokenizer, we can effectively represent network traffic data in a structured manner. By implementing this technique, we have achieved significant performance improvements compared to using text data with varying sizes;

   \item  We have designed a 15-layer BERT-based architecture with only $11$ million parameters for multi-category classification. We trained the model on PPFLE encoded data, which we refer to as \texttt{SecurityBERT};
  \item We evaluated the efficiency of our proposed approach using the Edge-IIoTset cyber security dataset~\cite{ferrag2022edge}. Various ML techniques have recently been tested on this dataset, providing a solid foundation for fair comparison. According to our experimental analysis, our method effectively identifies fourteen distinct types of attacks on an average CPU in less than $0.3$ seconds, achieving an overall accuracy of 98.2\%. To the best of our knowledge, this achievement showcases the highest accuracy ever attained among all ML algorithms, outperforming both the Convolutional Neural Network (CNN) and Transformer models.

\end{itemize}

This paper is organized as follows:
Section~\ref{sec:sec2} presents an exploration of the related work. Subsequently, Section~\ref{sec:sec3} outlines the significant steps in developing \texttt{SecurityBERT}. In Section~\ref{sec:sec4}, we evaluate the performance of the proposed model. Lastly, we conclude our research and provide insight into potential future research directions of interest in Section~\ref{sec:Conclusion}.

%---------------------------------------------
\section{Related work}
\label{sec:sec2}
%---------------------------------------------

As various researchers have already demonstrated, the BERT model proves to be an exceptional starting point for identifying cybersecurity threats.  BERT has been utilized in various fields, from detecting log anomalies to identifying malicious web requests.

A noteworthy study by Alkhatib \textit{et al.}~\cite{alkhatib2022can} demonstrated the feasibility of using BERT for learning the sequence of arbitration identifiers (IDs) in a Controller Area Network (CAN) via a ``masked language model'' unsupervised training objective. They proposed the CAN-BERT transformer model for anomaly detection in current automotive systems and showed that the BERT model outperforms its predecessors regarding accuracy and F1-score. Rahali \textit{et al.}~\cite{Rahali2021MalBERTMD}  introduced MalBERT, a tool that conducts static analysis on the source code of Android applications. They used BERT to comprehend the contextual relationships of code words and classify them into representative malware categories. Their results further underscored the high performance of transformer-based models in malicious software detection. 

Chen \textit{et al.}~\cite{Chen2022BERTLogAD} introduced BERT-Log, an anomaly detection and fault diagnosis approach in large-scale computer systems that treat system logs as natural language sequences. They leveraged a pre-trained BERT model to learn the semantic representation of normal and anomalous logs, fine-tuning the model with a fully connected neural network to detect abnormalities. Seyyar \textit{et al.}~\cite{seyyar2022attack} proposed a model for detecting anomalous HTTP requests in web applications, employing Deep Learning (DL) techniques and BERT. Aghaei \textit{et al.}~\cite{aghaei2022securebert} presented SecureBERT\footnote{While the names may sound similar, it is important to note that SecureBERT is separate from our recently introduced SecurityBERT.}, a language model tailored explicitly for cybersecurity tasks, focusing on Cyber Threat Intelligence (CTI) and automation. The SecureBERT model offers a practical way of transforming natural language CTI into machine-readable formats, thereby minimizing the necessity for labor-intensive manual analysis. The authors devised a unique tokenizer and a method for adjusting pre-trained weights to ensure that SecureBERT understands general English and cybersecurity-related text. However, SecureBERT is not designed to process network-based cyber threat attacks. 

CyBERT, introduced by Ranade et al.~\cite{Ranade}, is a custom version of BERT designed specifically for cybersecurity applications. This model has been fine-tuned using a vast corpus of cybersecurity data to enhance its ability to process intricate information concerning threats, attacks, and vulnerabilities.
K. Yu et al.~\cite{9627827} explored a deep-learning-based approach for detecting advanced persistent threats (APTs) in the Industrial Internet of Things (IIoT), using the BERT model to address the challenges of long attack sequences. Their experimental results demonstrate the method's effectiveness, yielding high accuracy and a low false alarm rate in APT detection. B. Breve et al.~\cite{9953113} proposed using NLP techniques, specifically a BERT-based model, to detect potentially harmful automation rules in trigger-action IoT platforms that could breach user security or privacy. Their evaluation on the If-This-Then-That platform with over 76,000 rules demonstrated high accuracy, significantly outperforming traditional information flow analysis methods. Recently Z. Wang et al.~\cite{WANG2024122045} developed BERT-of-Theseus, Vision Transformer, and PoolFormer (BT-TPF), an IoT intrusion detection model tailored for resource-limited IoT environments, using a knowledge-distillation approach. The model employs a Siamese network for feature reduction and a Vision Transformer to train a compact Poolformer model, achieving a significant parameter reduction while maintaining high accuracy. The aforementioned studies leverage pre-trained BERT models and customize them to meet their unique security needs by fine-tuning or using them as feature generators. These models benefit from the textual form and sequential nature of their security-related data, including sources such as code, emails, and log sequences. These studies effectively utilize BERT's ability to comprehend contextual relationships within sequences to carry out precise detection and classification tasks.

In cyber threat detection, it is vital to compare different research efforts. In real-world cyber threat detection scenarios, support is crucial for extracting features from network traffic, often relying on PCAP files. In addition to analyzing real packet data and detecting cyber threats on networks, it is important to consider privacy in training data, especially since IoT devices and network data may contain sensitive information. Given the uniqueness of each network infrastructure and the need for high accuracy through fine-tuning or new training, sharing actual network traffic data for training purposes can raise privacy concerns. \texttt{SecurityBERT} has been developed as a pioneering, lightweight, privacy-preserving architecture specifically designed with this consideration in mind. TABLE~\ref{tab:comp1} provides a comparison of various recent works on cyber threat detection in terms of four key parameters: 
\begin{itemize}
    \item \textbf{D = Detect:} Network-based Cyber Threat Detection
    \item \textbf{L = LLM:} Utilization of LLMs
    \item \textbf{N = Network PCAP:} Packet data analysis of a traffic
        \item \textbf{P = Privacy:} Privacy-preserving training data
\end{itemize}

\begin{table}[ht!]
\centering
\caption{Comparison with recent works on cyber threat detection}\label{tab:comp1}
\begin{tabular}{|l|l|l|l|l|l|}
\hline
\rowcolor{lightgray}\textbf{Frameworks} & \textbf{Year} & \textbf{D} & \textbf{L} &  \textbf{N} & \textbf{P}\\ \hline
Alkhatib \textit{et al.}~\cite{alkhatib2022can}  &         2022          &  \ding{55}      &   \cmark   &   \ding{55} &  \ding{55}\\ \hline
Rahali \textit{et al.}~\cite{Rahali2021MalBERTMD} &         2022      &   \ding{55}     &  \cmark         &   \ding{55}&  \ding{55} \\ \hline
Aghaei \textit{et al.} \cite{aghaei2022securebert}             &         2022      &   \ding{55}     &  \cmark         &   \ding{55}&  \ding{55} \\ \hline
Thapa \textit{et al.} \cite{thapa2022transformer} &         2022      &   \ding{55}     &  \cmark          &   \ding{55}&  \ding{55} \\ \hline
  Hamouda \textit{et al.} \cite{hamouda2022ppss}             &         2022      &   \cmark     &  \ding{55}      &     \cmark  &  \ding{55}   \\ \hline
Friha \textit{et al.} \cite{friha2022felids}               &             2022  &   \cmark     &  \ding{55}      &   \cmark   &  \ding{55} \\ \hline

Chen \textit{et al.}~\cite{Chen2022BERTLogAD}  &         2022      &   \ding{55}     &  \cmark         &   \ding{55} &  \ding{55} \\ \hline
Seyyar \textit{et al.}~\cite{seyyar2022attack} &         2022      &   \ding{55}     &  \cmark      &   \ding{55} &  \ding{55} \\ \hline
Selvaraja \textit{et al.} \cite{selvarajan2023artificial}               &    2023   &   \cmark     &  \ding{55}       &  \cmark    &  \ding{55}   \\ \hline

B. Breve \textit{et al.} \cite{9953113}               &    2023   &   \ding{55}      &  \cmark        & \ding{55}     &  \ding{55}  \\ \hline

Chen \textit{et al.} \cite{chen2023diversevul}             &         2023     &   \ding{55}    &  \cmark      &   \ding{55}   &  \ding{55}  \\ \hline
Douiba \textit{et al.} \cite{douiba2023improved}
               &  2023   &   \cmark     &  \ding{55}       &    \cmark   &  \ding{55} \\ \hline
 Jahangir \textit{et al.} \cite{jahangir2023deep}              &       2023        &   \cmark     &  \ding{55}       &   \cmark   &  \ding{55}  \\ \hline
Hu \textit{et al.} \cite{hu2023towards}               &    2023           &   \cmark     &  \ding{55}       &   \cmark   &  \ding{55} \\ \hline
K. Yu \textit{et al.}~\cite{9627827}              &       2023        &   \cmark     &    \cmark      &   \ding{55}   &  \ding{55}  \\ \hline
Hu \textit{et al.} \cite{hu2023towards}               &    2023           &   \cmark     &  \ding{55}       &   \cmark  &  \ding{55}  \\ \hline

Friha \textit{et al.} \cite{friha20232df}               &             2023  &   \cmark     &  \ding{55}      &    \cmark   &  \ding{55} \\ \hline
Chakraborty \textit{et al.} \cite{chakraborty2023intelligent}               &     2023          &   \cmark     &  \ding{55}      &    \cmark    &  \ding{55}  \\ \hline
Wang \textit{et al.} \cite{wang2023resilient}               &         2023      &   \cmark     &  \ding{55}      &   \cmark   &  \ding{55}  \\ \hline
Liu \textit{et al.} \cite{liu2023cps}               &    2023           &   \cmark     &  \ding{55}      &    \cmark   &  \ding{55} \\ \hline
Aouedi \textit{et al.} \cite{aouedi2023f}               &            2023   &   \cmark     &  \ding{55}      &    \cmark  &  \ding{55}   \\ \hline
Z. Wang \textit{et al.} \cite{WANG2024122045}               &            2023   &   \cmark     &  \cmark      &    \cmark  &  \ding{55}   \\ \hline
\textbf{SecurityBERT}  &   2023     &    \large\cmark    &   \large\cmark    &    \large\cmark &  \large\cmark \\ \hline
\end{tabular}\\

\vspace{5pt}
\checkmark : Supported, \ding{55}: Not Supported.
\end{table}

The majority of the research conducted in 2022, including \cite{Rahali2021MalBERTMD}, \cite{aghaei2022securebert}, and \cite{alkhatib2022can}, integrated LLMs, but they did not support detection, nor did they utilize packet data. However, contrary to the norm, Hamouda et al.~\cite{hamouda2022ppss} (2022) and Friha et al.~\cite{friha2022felids} (2022) demonstrated support for cyber threat detection and utilized packet data but did not rely on the capabilities of LLMs. Works from 2023, such as \cite{selvarajan2023artificial}, \cite{douiba2023improved}, and \cite{wang2023resilient}, emphasize more cyber threat detection and the use of packet data, but they largely lack in the application of LLMs.

%-----------------------------------------------------
\section{SecurityBERT architecture design}
\label{sec:sec3}
%-----------------------------------------------------

\begin{figure*}[htp]
    \centering
    \includegraphics[width=1\textwidth]{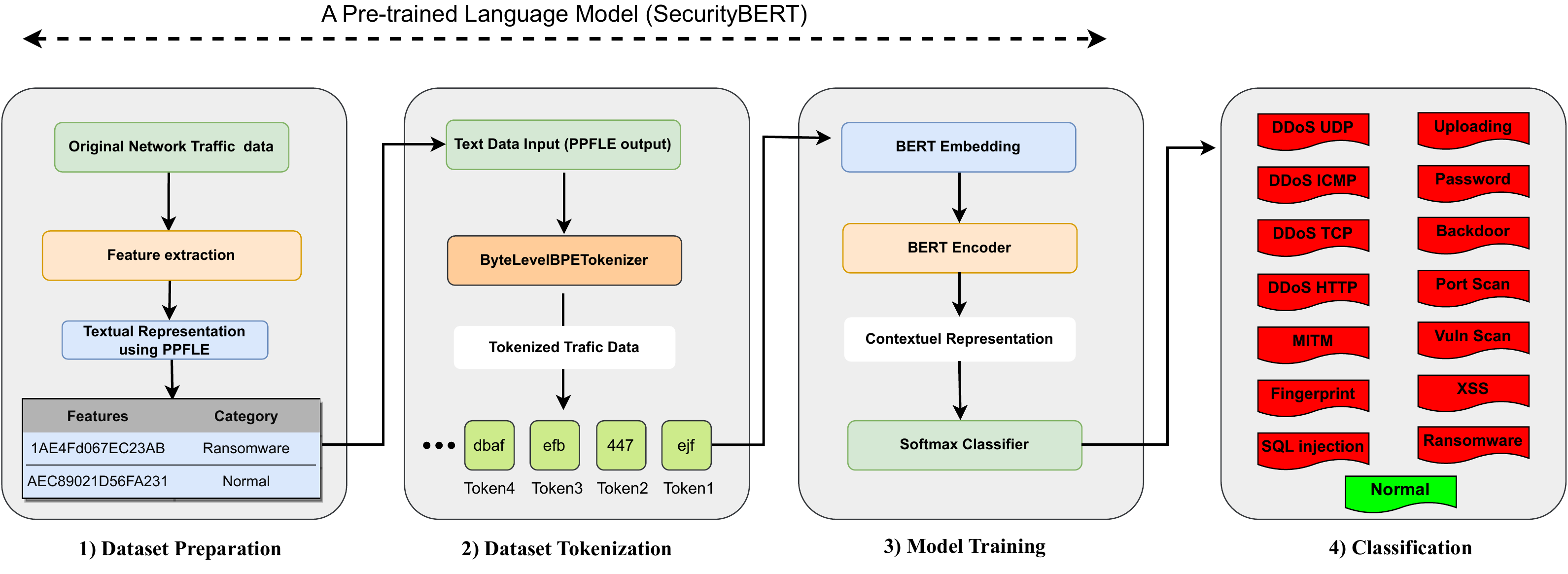}
  \caption{High-level workflow of our SecurityBERT model.}
    \label{fig:workflow}
\end{figure*}

FIGURE~\ref{fig:workflow} visually presents the comprehensive workflow of the model, encompassing all relevant steps from dataset preparation to classification. Each of these steps will be extensively covered in this section. Developing a BERT model from the ground up for network-based cyber threat detection demands a thorough and intricate approach. Below is a comprehensive outline detailing the main steps in the process:

\begin{steps}
\caption{Main steps of building SecurityBERT}

\begin{algorithmic}[1]
\State \textbf{Dataset Utilization}

\State \textbf{Feature Extraction}

\State \textbf{Privacy-Preserving Fixed-Length Encoding (PPFLE)}
\State \textbf{Byte-level BPE (BBPE) Tokenizer}

\State \textbf{SecurityBERT Embedding}

\State \textbf{Contextual Representation}

\State \textbf{Training SecurityBERT}
\begin{itemize}
    \item Text Normalization
    \item Text Tokenization
    \item Frequency Filtering
    \item Vocabulary Creation
    \item Special Token Addition
    \item Tokenizer Training
\end{itemize}

\State \textbf{Fine-tuning with Softmax activation function}

\end{algorithmic}
\end{steps}

%-----------------------------------------------------------
\subsection{Dataset Utilization (Edge-IIoTset Dataset)} 
%-----------------------------------------------------------

Generating our dataset through real-life traffic analysis would be time-consuming, and there's the risk of specific attacks not being adequately simulated or missing from our dataset. Hence, acquiring and utilizing realistic datasets for our research is crucial.

\begin{figure}[htbp]
    \centering
    \includegraphics[width=0.5\textwidth]{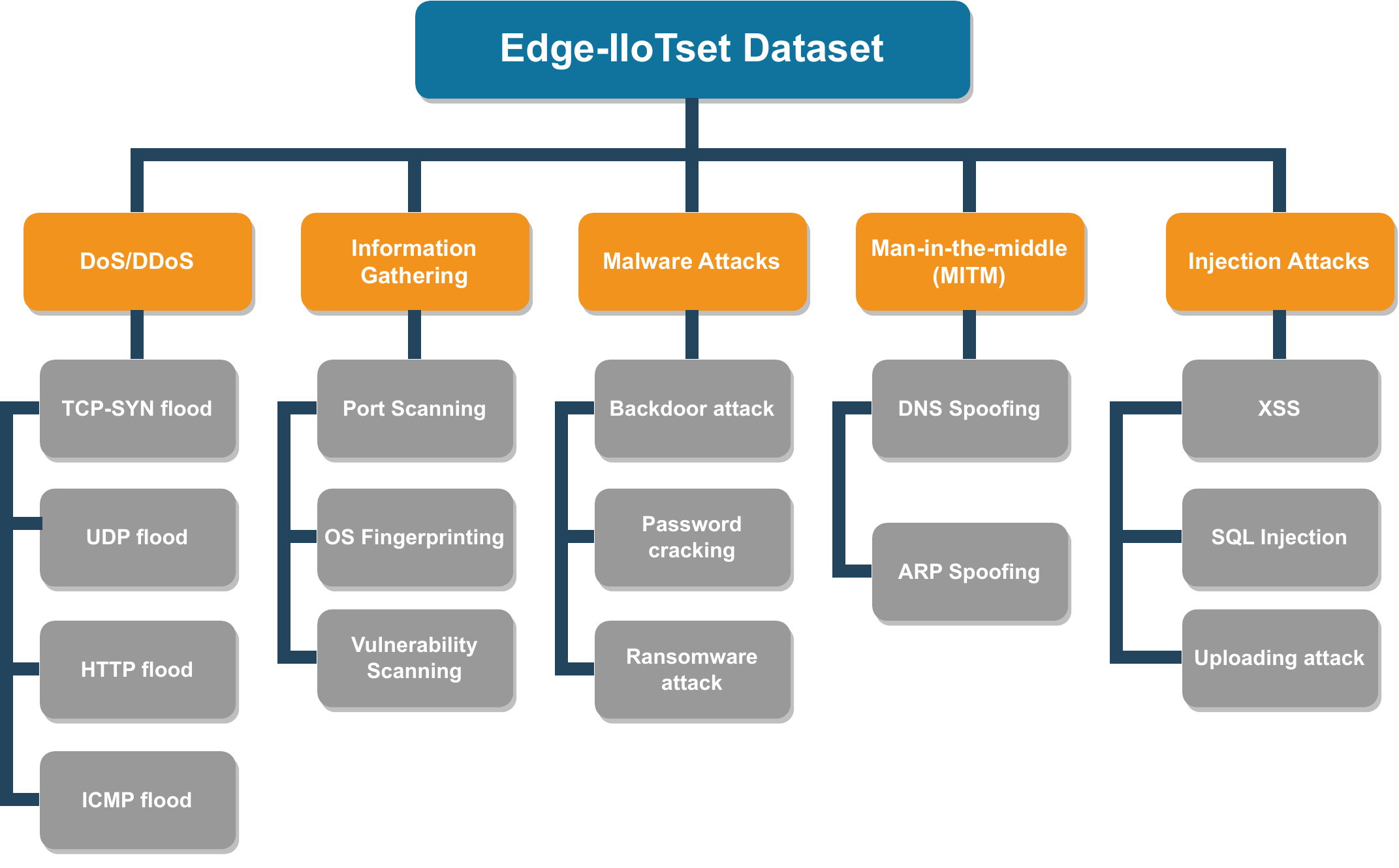}
  \caption{Categories of the Edge-IIoTset dataset}
    \label{fig:EDGE_dataset}
\end{figure}

Cybersecurity and network security data can be gathered from various online sources using open-source databases and repositories. Notable examples include the Common Vulnerabilities and Exposures (CVE) database, the Open Web Application Security Project (OWASP), and numerous others for network security~\cite{ferrag2020deep}. The primary challenge presented by these sources is their heavy reliance on artificial scenarios, which results in a deficiency of authentic data. Training a model exclusively on such data can potentially lead to unrealistic outcomes. Furthermore, most of these databases do not include packet network data, which poses a challenge in simulating realistic scenarios.
Our primary aim is to opt for a dataset that tackles this constraint by strongly emphasizing genuine network data. Furthermore, we intend to ensure maximum diversity within this dataset, encompassing a comprehensive range of attack types, including ransomware, XSS, SQL injection, DoS, and other widely recognized attack categories. This diversified dataset's rationale is to assess our newly proposed model's classification capabilities comprehensively. In 2022, Ferrag et al. introduced Edge-IIoTset~\cite{ferrag2022edge}, a new and extensive cybersecurity dataset specifically designed for IoT and IIoT applications. This dataset serves as a valuable resource for ML-based intrusion detection systems. The Edge-IIoTset dataset includes diverse devices, sensors, protocols, and cloud/edge configurations, rendering it highly representative of real-world scenarios and aligning perfectly with our research objectives. 
This dataset contains fifteen ($15$) attacks related to the Internet of Things (IoT) and Industrial IoT (IIoT) connectivity protocols, categorized into five threats: DoS/DDoS attacks, Information gathering, Man-in-the-middle (MITM), Injection attacks, and Malware attacks, which can be seen in Figure~\ref{fig:EDGE_dataset}. The DoS/DDoS attack category encompasses TCP SYN Flood, UDP flood, HTTP flood, and ICMP flood attacks. The Information Gathering category includes attacks like port scanning, operating system fingerprinting, and vulnerability scanning. MITM attacks include tactics such as DNS Spoofing and ARP Spoofing. Injection attacks include Cross-Site Scripting (XSS), SQL injection, and file-uploading attacks. Lastly, the Malware category covers backdoors, password crackers, and ransomware attacks.
%-----------------------------------------------------
\subsection{Features extraction} 
%-----------------------------------------------------
Given a PCAP file with a network traffic log, we extract relevant features from a specific time window and return them in a structured format suitable for analysis. Specifically, we identify and separate each network flow in the PCAP file. For each flow identified, we extract a set of predefined features. Then, we organize the extracted features into a CSV file format for analysis. We removed null features from the Edge-IIoTset dataset during our initial exploration, identifying $61$ distinct and diverse features. These features are sufficiently various to distinguish the distinctive patterns of network attacks exclusively.

The Edge-IIoTset dataset comprises features gathered from various sources, including network traffic, logs, system resources, and alerts. To better understand these features, the initial 15 can be seen in TABLE~\ref{tab:features}. For a comprehensive view of all 61 features, please see Table 7 in~\cite{ferrag2022edge}.
\begin{table}[ht!]

    \centering
    \begin{tabular}{|c|l|c|l|}
 \hline
        \rowcolor{lightgray}\textbf{N°} & \textbf{Name} & \textbf{Prot. Layer} & \textbf{Type} \\
        \hline
        1 & frame.time & Frame & Date and time \\ \hline
        2 & ip.src\_host & IP & Character string \\ \hline
        3 & ip.dst\_host & IP & Character string \\ \hline
        4 & arp.dst.proto\_ipv4 & ARP & IPv4 address \\ \hline
        5 & arp.opcode & ARP & Unsigned integer \\ \hline
        6 & arp.hw.size & ARP & Unsigned integer \\ \hline
        7 & arp.src.proto\_ipv4 & ARP & IPv4 address \\ \hline
        8 & icmp.checksum & ICMP & Unsigned integer \\ \hline
        9 & icmp.seq\_le & ICMP & Unsigned integer \\ \hline
        10 & icmp.transmit\_timestamp & ICMP & Unsigned integer  \\ \hline
        11 & icmp.unused & ICMP & Sequence of bytes \\ \hline
        12 & http.file\_data & HTTP & Character string \\ \hline
        13 & http.content\_length & HTTP & Unsigned integer \\ \hline
        14 & http.request.uri.query & HTTP & Character string \\ \hline
        15 & http.request.method & HTTP & Character string \\ \hline
        . & $\cdots$ & $\cdots$ & $\cdots$ \\ \hline
        . & $\cdots$ & $\cdots$ & $\cdots$ \\ \hline
        61 & mbtcp.unit\_id & Modbus/TCP & Unsigned Integer \\ \hline
    \end{tabular}
    \caption{The first 15 features gathered from PCAP files.}
   \label{tab:features}
\end{table}

Numerous studies have already shown that these 61 distinct features are sufficient to detect specific network-based cyberattacks. This dataset, therefore, serves as an optimal foundation for comparing various ML algorithms~\cite{Transformer95Ferrag}. After discussing the exact architectural design, the evaluation and comparison of \texttt{SecurityBERT} with other research will be detailed in Section~\ref{sec:sec4}.

\subsection{Privacy-Preserving Fixed-Length Encoding}
A pivotal aspect of the design involves representing the unstructured network data in a manner that allows BERT to comprehend the context and relationships between various features. 
BERT is designed to understand English proficiently but may not be the most suitable ML model for comprehending relationships between numbers. In our case, many features are numerical values, i.e., unsigned integers, not strings (as illustrated in TABLE~\ref{tab:features}), making it difficult to discern their interrelationships using natural language processing methods.

To leverage the power of BERT natural language understanding, we process the dataset, comprising numerical and categorical values, and transform it into textual representation. Specifically, we added context to the data by incorporating column names and concatenating them with their respective values. Then, each new value is hashed and combined with other hashed values within the same instance, resulting in the generation of a sequence.  By employing this technique, we have developed a new language comprehensible to BERT and introduced privacy into the training data through cryptographic hash functions.
We call this novel textual representation technique as \textit{Privacy-Preserving
Fixed-Length Encoding (PPFLE)}. 

Significant similarities in log files, TCP scans, and memory dumps may lead to misinterpretation and incorrect classification of various attacks. Employing a hash function allows for handling even minor deviations in the data, effectively representing them as distinct data points for ML. Moreover, specific attacks, like UDP scans and others that are challenging to represent as plain text, can be better understood by the model when they are converted into hashed values. Put simply,  we have developed a new linguistic format that the BERT model comprehends much more effectively than mere numerical data, and it aligns more closely with the natural English language for which the BERT model is specifically tailored. 

Through this method, we fashioned a representation of the numbers that closely align with natural language, allowing the model to attain enhanced classification accuracy, as detailed in Section~\ref{sec:experimental}. Correctly converting network data and applying PPFLE can achieve higher accuracy than using the original pre-trained BERT model architecture.

%--------------------------------------------------
\subsubsection*{PPFLE description}
%--------------------------------------------------

The objectives of PPFLE are twofold. On the one hand, it is designed to convert unstructured network data into a structured format that better mimics the natural English language, aligning well with the BERT model's specialization. On the other hand, it focuses on maintaining privacy by ensuring that only encoded data is observed, thereby hiding sensitive information in the network data while preserving key classification features.

Let us define a matrix denoted by $\mathbf{M}$ with $i$ rows and $j$ columns. Here, $\mathbf{M}[i, j]$ represents the matrix element at the intersection of the ith row and jth column in $\mathbf{M}$. We denote the ith row of $\mathbf{M}$ by $r_i=\mathbf{M}[i, :]$. In $\mathbf{M}$, the first row contains the column names, which serve as labels or identifiers for each column.  We denote these column names as $c_j$, where $j$ represents the column index, i.e., $\mathbf{M}[1,j] = c_j$.

Let us define $s(i, j)$ as a concatenation operation where the column name $c_j$, a dollar sign, and the value of the  $j$th  column in the $(i+1)$th row $r_i$ are concatenated into a single string,  excluding the first row which contains the column names, i.e.,
\begin{equation}
s(i, j) = c_j \; \parallel \; \text{"\$"} \; \parallel \; \mathbf{M}[i+1,j]
\end{equation}
Next, define $H(x)$ as a hashing operation on a string $x$ and let 
$L$ is a list where each element is separated by a space, i.e., $L = \{l_1 \, l_2 \, l_3 \, \ldots \, l_k\}$.

Then, the textual representation of each row $i$ in the matrix $\mathbf{M}$  can be expressed as follows:
\begin{equation}
L \leftarrow L \cup \{H(s(i, n))\} \,\,\,  \forall \; (1 \leq n \leq  j
)\end{equation}
Repeating this procedure for each row in $\mathbf{M}$, we obtain a new matrix called DataList denoted as $\mathcal{DL}$. In $\mathcal{DL}$, each row represents an $L$ list, i.e.;

\begin{equation*}
\mathcal{DL}  = \left[ {\begin{array}{ccccccccccccc}
L_1=[H(s(1,1)) \, H(s(1,2)) \dots \,H(s(1,j))] \\
L_2=[H(s(2,1)) \, H(s(2,2)) \dots \,H(s(2,j))] \\
 \vdots \\ 
L_i=[H(s(i,1)) \, H(s(i,2)) \dots \, H(s(i,j))] \\
\end{array} } \right] 
\end{equation*}

\begin{figure}[htbp]
    \centering
    \includegraphics[width=0.35\textwidth]{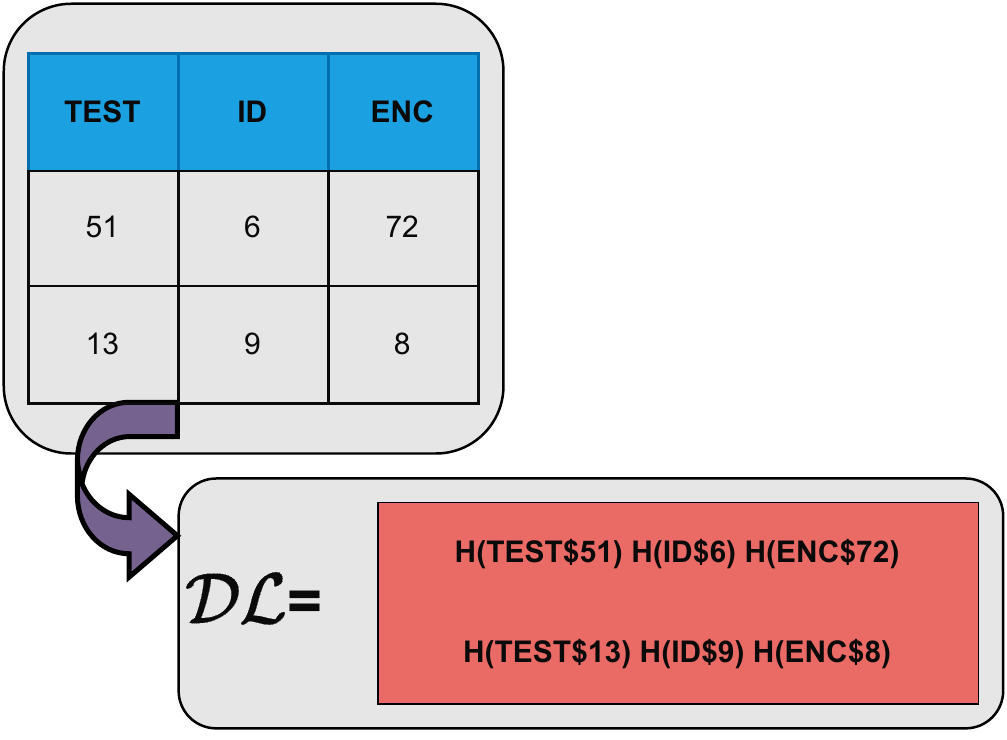}
  \caption{Creating DataList example.}
    \label{fig:DataListcreate}
\end{figure}

In other words, the DataList $\mathcal{DL}$ lists where each inner list contains the hashed, concatenated column values for each row in the matrix $\mathbf{M}$. The $\mathbf{M}$ matrix in ML is commonly called a DataFrame. FIGURE~\ref{fig:DataListcreate} showcases a simple DataList creation example. The pseudocode of the PPFLE algorithm can be seen in Algorithm~\ref{alg:algo1}. 

\begin{algorithm}
\caption{Privacy-Preserving Fixed-Length Encoding}\label{alg:algo1}
\begin{algorithmic}[1]
\Require Matrix $\mathbf{M}$ with $i$ rows and $j$ columns
\Procedure{PPFLE}{$\mathbf{M}$}
\State $\mathcal{DL} \gets []$  \Comment{Initialize $\mathcal{DL}$ to be empty}
\For{$m = 1$ to $i$} \Comment{Iterate through rows in $\mathbf{M}$}
\State $L= []$ \Comment{Initialize $L$ to be an empty list}
\For{$n = 1$ to $j$} \Comment{Column iteration}
\State $ L \gets H(s(m, n))$ \Comment{Append $H(x)$ to $L$}
\EndFor

\State $ \mathcal{DL} \gets L$ \Comment{Append $L$ to $\mathcal{DL}$}
\EndFor
\State \Return $\mathcal{DL}$
\EndProcedure
\end{algorithmic}
\end{algorithm}
The PPFLE algorithm, despite its simplicity, effectively translates unstructured data into a fixed-length format. This representation mirrors the characteristics of natural languages, offering considerable advantages when utilized by ML algorithms. 

\subsubsection*{Removed features for PPFLE encoding}
%-----------------------------------------------------------

A natural question arises as to whether all $61$ features are suitable for PPFLE encoding. For instance, features like "ip.src\_host" and "ip.dst\_host" contain IP addresses, which can lead to overfitting, especially if they have unique identifiers or particular details that don't generalize well in different network. Similarly, hashing timestamps with millisecond precision could introduce confusion during training, so it may be necessary to remove such features if one intends to apply PPFLE encoding. For this reason, several features related to network traffic and packet captures were excluded.  High-cardinality features such as "http.request.full\_uri" can be challenging to encode effectively and might not offer generalizable patterns. Features with potential redundancy, like the presence of both "ip.src\_host" and "arp.src.proto\_ipv4", could introduce multicollinearity, affecting model stability. Features such as "frame.time", indicating packet capture timestamps, might not directly relate to the predictive modeling task. Other columns like "tcp.payload" and "http.file\_data" represent raw data payloads, which, without extensive preprocessing, could introduce noise rather than clarity. Removing these columns streamlines the dataset, enhancing computational efficiency and ensuring the model focuses on the most relevant and generalizable patterns while maintaining user privacy.

%-------------------------------------------------
\subsection{Byte-level BPE (BBPE) Tokenizer}
%-------------------------------------------------

Tokenization is performed on the PPFLE-encoded data. This ensures that no sensitive information is fed to the model during training. 
A natural question arises: Doesn't the PPFLE compromise the semantics of network data, rendering tokenization unfeasible? By applying PPFLE encoding, we convert numerical values to align with the characteristics of natural language more closely. Each feature is encoded independently, allowing the adjacent hashed values to provide the model with sufficient information about the type of attacks it encounters. Hashing all 61 features together, however, would destroy the semantics of the attacks. 

\begin{figure}[htbp]
    \centering
    \includegraphics[width=0.5\textwidth]{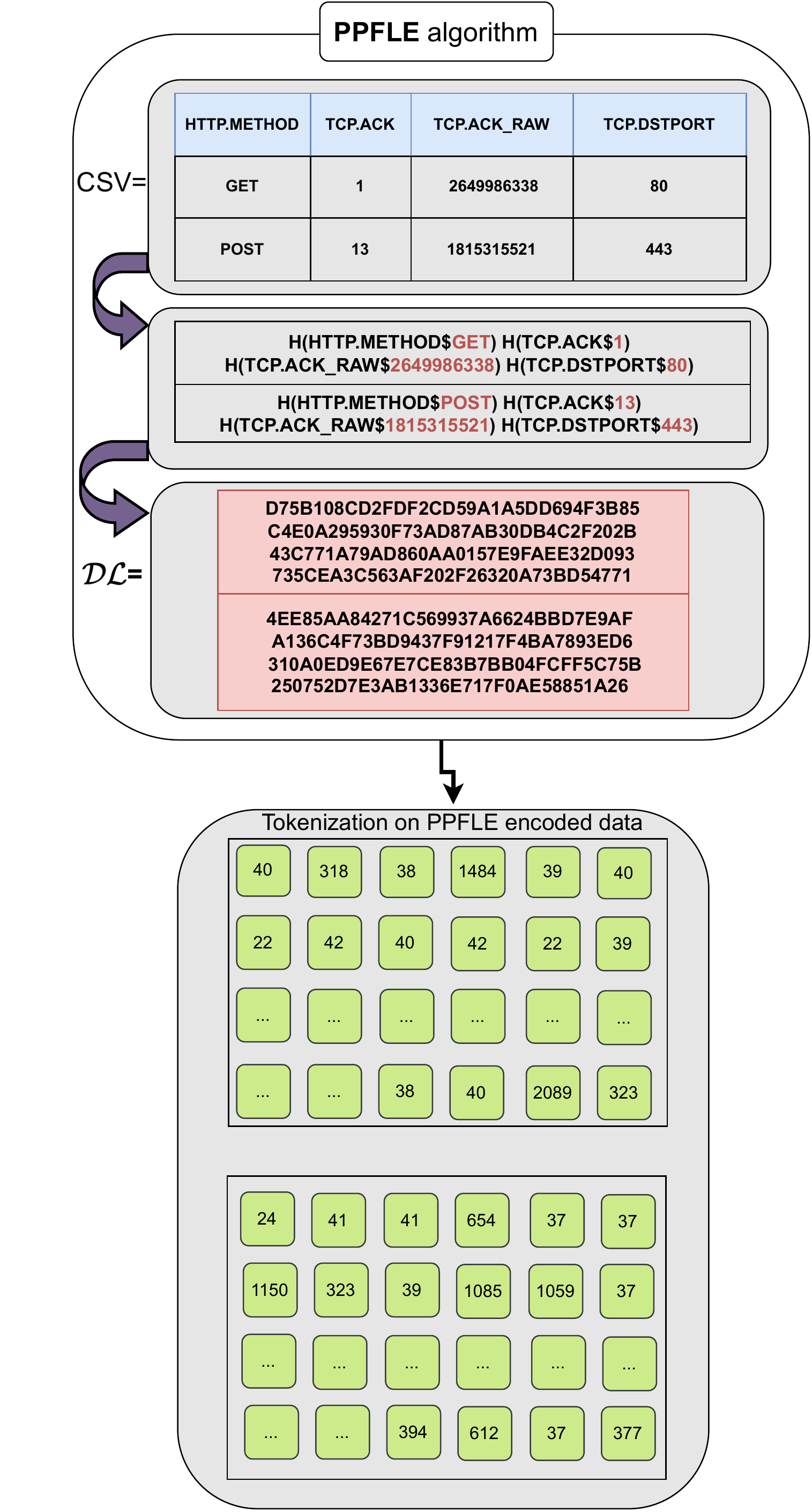}
  \caption{PPFLE encoding and BBPE tokenization example}
    \label{fig:PPFLE_example}
\end{figure}

Figure~\ref{fig:PPFLE_example} demonstrates the functionality of the PPFLE algorithm, including tokenization.

For instance, using PPFLE to encode a feature for an attack on port 443 with the GET method would appear as: \texttt{H(TCP.DSTPORT\$443) H(HTTP.METHOD\$GET)}. Conversely, a DNS poisoning attack would have a distinct representation, lacking any \texttt{HTTP.METHOD} and thus consistently hashed as \texttt{H(HTTP.METHOD\$0)}.
It has turned out during our experimental analysis that these 61 features are highly effective in representing different types of network attacks with great accuracy. Furthermore, the model can recognize all attack patterns based on these features, even when hashed.
For the data encoded with PPFLE, we employed the \texttt{ByteLevelBPETokenizer} from the Hugging Face Transformers library. This tokenizer, initially utilized for GPT-2~\cite{wolf2019huggingface}, breaks down text into subword units for tokenization. It is based on the Byte-Pair Encoding (BPE) algorithm~\cite{shibata1999byte}, a data compression technique that replaces the most frequent pair of consecutive bytes in a sequence with a single, unused byte. The \texttt{ByteLevelBPETokenizer} is particularly useful for handling out-of-vocabulary (OOV) words, which are not present in the tokenizer's vocabulary of human language~\cite{araabi2022effective}. By breaking down our language presentation of network traffic data into smaller subwords likely present in the tokenizer's vocabulary as a sequence of bytes, we can efficiently process traffic data by leveraging the power of BERT.

During the training of the tokenizer, a vocabulary size of 5000 was employed, along with a set of specific tokens, including \texttt{["<s>", "<pad>", "</s>", "<unk>", "<mask>"]}. The tokenizer's training involved utilizing the file name, setting the vocabulary size, establishing a minimum frequency of 2, and incorporating the list of special tokens. For a visual representation of the various tokens within the PPFLE encoded data, refer to FIGURE~\ref{fig:PPFLE_example}. Understanding the semantics of these tokens functions similarly to interpreting a typical sentence. The hash output for a specific attack remains constant; thus, if these subword hexadecimal values appear in a particular sequence, BERT can recognize that this unique order corresponds to a hash output and categorize it as a specific attack.

%------------------------------------------------
\subsection{SecurityBERT embedding} 
%------------------------------------------------

Algorithm \ref{alg:algo3} showcases the \texttt{SecurityBERT} embedding. The algorithm starts by setting the $chunk\_size$ to $5000$.

\begin{algorithm}
\caption{Encode Evaluation Data Sequences}\label{alg:algo3}
\begin{algorithmic}[1]
\State $chunk\_size$ $\gets 5000$
\State $num\_chunks \gets \lceil$ \texttt{len}($eval\_data$) / $chunk\_size \rceil$
\State $input\_ids\_eval$ $\gets []$ \Comment{Initialize as an empty list} 
\State $attention\_masks\_eval$ $\gets []$ \Comment{Initialize as an empty list} 
\For{$i$ = $0$ to $num\_chunks$}
    \State $start\_idx \gets i \times chunk\_size$
    \State $end\_idx \gets (i+1) \times chunk\_size$
    \State $chunk \gets eval\_data[start\_idx:end\_idx]$
    \State $encoded\_seqs \gets$ encode($chunk$)  
 
    \State $iic,amc \gets$ UNPACK($encoded\_seqs$)

    \State \textbf{append} $icc$ to $input\_ids\_eval$
    
    \State \textbf{append} $amc$ to $attention\_masks\_eval$
\EndFor
\State \textbf{concatenate} the input IDs and attention masks as $input\_ids\_eval$, $attention\_masks\_eval$
\end{algorithmic}
\end{algorithm}

 It then calculates the number of chunks, $num\_chunks$, by dividing the length of the $eval\_data$ by $chunk\_size$, and rounding up to the nearest integer. Two empty lists, $input\_ids\_eval$ and $attention\_masks\_eval$, are initialized to hold the encoded input IDs and attention masks, respectively. The algorithm then enters a loop, iterating from $0$ to \texttt{num\_chunks}. This loop determines the start and end indices for each chunk of the \texttt{eval\_data}. It retrieves a chunk of data using these indices and encodes each sequence in the chunk, storing the result in \texttt{encoded\_seqs}. This encoded data is then unpacked into two components: \texttt{input\_ids\_chunk} and \texttt{attention\_masks\_chunk}, denoted by $iic$ and $amc$. These components are appended to the \texttt{input\_ids\_eval} and \texttt{attention\_masks\_eval} lists. Once all chunks have been processed, the algorithm concatenates all the input IDs and attention masks in \texttt{input\_ids\_eval} and \texttt{attention\_masks\_eval} respectively, along dimension $0$, thereby creating a complete set of input IDs and attention masks for the evaluation data. Here, we note that the \texttt{input\_ids\_eval} and \texttt{attention\_masks\_eval} are important components of the input to transformer-based models. The  \texttt{input\_ids\_eval} is a sequence of integers representing the input data after being tokenized. Each integer maps to a token in the model's vocabulary.

The \texttt{attention\_masks\_eval} informs the model about which tokens should be attended to and which should not. In many cases, sequences are padded with special tokens to make all sequences the same length for batch processing. Attention masks prevent the model from attending to these padding tokens. Typically, an attention mask has the same length as the corresponding \texttt{input\_ids} sequence and contains a $1$ for real tokens and a $0$ for padding tokens.
%

%-----------------------------------------------------------
\subsection{Contextual representation}
%-----------------------------------------------------------

We adopted the BERT architecture, which leverages transformers for textual representation and cyber threat classification. Specifically, we pre-trained our \texttt{SecurityBERT} using our newly created tokenized dataset. In this process, \texttt{SecurityBERT} takes each token from the tokenized text and represents it as an embedding vector, denoted as $X \in R^d$, where $d$ represents the dimensionality of the embedding space. Then \texttt{SecurityBERT} utilizes a transformer-based architecture consisting of multiple encoder layers. Each encoder layer comprises multi-head self-attention mechanisms and position-wise feed-forward neural networks. The self-attention mechanism~\cite{vaswani2017attention} allows the model to capture dependencies and relationships between words within a sentence, thus facilitating contextual understanding. The self-attention mechanism in BERT can be mathematically expressed as follows:
\begin{equation}
\text{Attention(Q, K, V)} = \sigma\left(\frac{QK^T}{\sqrt{d_k}}\right)V
\end{equation}

\noindent, where $\sigma$ is the softmax function, $Q$, $K$, and $V$ are the query, key, and value matrices, respectively, $d_k$ represents the dimensionality of the keys vector, and $T$ denotes the transpose operation. Through self-attention, BERT encodes contextual representations by capturing the importance of different words within a sentence based on their semantic and syntactic relationships. The resulting contextual embeddings are obtained through feed-forward operations and layer normalization. 

%----------------------------------------------------
\subsection{Training SecurityBERT}
%----------------------------------------------------

The training of \texttt{SecurityBERT} involves several crucial steps, each carefully calibrated to ensure optimal performance in security-centric tasks. These steps include data collection and preprocessing, tokenizer training, model configuration, and the training process itself. \texttt{SecurityBERT} works with PPFLE-encoded data, simplifying certain steps in the tokenizer training process and requiring alternative approaches for other steps.  Here, we detail the distinct aspects of \texttt{SecurityBERT's} training process.

%-------------------------------------------------
\subsubsection{Text Normalization}
%-------------------------------------------------

\begin{equation}
n(D) = \{n(d) | d \in D\}
\end{equation}
In this function, $n(D)$ represents the normalization process applied to each document $d$ in the set of all documents $D$. Text normalization typically involves converting all text to lowercase, removing punctuation, and sometimes even stemming or lemmatizing words (reducing them to their root form). This process is part of the original BERT architecture; however, when working with PPFLE-encoded data, this element becomes unnecessary and does not provide any extra value to our architecture.

\begin{table*}[ht]
\centering

\begin{tabular}{|p{1.5in}|p{1.3in}|p{1.3in}|p{1.3in}|}
\hline
\rowcolor{lightgray}
\textbf{Feature} & \textbf{BertBase} & \textbf{BertLarge} & \textbf{SecurityBERT} \\ \hline
Model Type & Pretrained model on English language & Pretrained model on English language & Pretrained model on Network Data \\ \hline
Word Embeddings Size & 30522 x 768 & 30522 x 1024 & 30522 x 256 \\ \hline
Position Embeddings Size & 512 x 768 & 512 x 1024 & 512 x 256 \\ \hline
Token Type Embeddings Size & 2 x 768 & 2 x 1024 & 2 x 256 \\ \hline
Number of Layers in Encoder & 12 & 24 & 4 \\ \hline
Size of Query, Key, Value in Attention & 768 & 1024 & 256 \\ \hline
Size of Intermediate Layer in Transformer & 3072 & 4096 & 1024 \\ \hline
Output Size of Transformer Layer & 768 & 1024 & 256 \\ \hline
Pooler Output Size & 768 & 1024 & 256 \\ \hline
Number of Parameters & ~110M & ~340M & ~11M \\ \hline
Additional Components & None (General-purpose model) & None (General-purpose model) & Dropout (for regularization) and Classifier (linear layer for classification tasks) \\ \hline
\end{tabular}
\caption{Comparison of Original BertBase, BertLarge, and SecurityBERT}
\label{tab:bert_comparison}
\end{table*}

\subsubsection{Tokenization}
\begin{equation}
t(d) = \{t(w) | w \in d, d \in D\}
\end{equation}
The tokenization function $t(d)$ breaks down each document $d$ in the set $D$ into its constituent words or tokens $w$. These tokens are the basic units of text that a machine-learning model can understand and process.
\subsubsection{Frequency Filtering}
\begin{equation}
f(D, F) = \{w \in D | \text{freq}(w, D) \geq F\}
\end{equation}
This function $f(D, F)$ defines a High-Pass filter, cutting out tokens $w$ that have a frequency of occurrence, $\text{freq}(w, D)$ less than the minimum frequency $F$ in the set of all documents $D$. This is to remove rare words that might not provide much informational value for further processing or model training.
\subsubsection{Vocabulary Creation}
\begin{equation}
v(D, V) = \{w | w \in D, \text{rank}(w, D) \leq V\}
\end{equation}
Here, the function $v(D, V)$ creates a vocabulary by choosing the top $V$ words $w$ from the set of all documents $D$ based on their frequency rank $\text{rank}(w, D)$. This forms the vocabulary that the model will recognize.
\subsubsection{Special Token Addition}

\begin{equation}
v' = v \cup S
\end{equation}

This states that the new vocabulary, $v'$, is a union of the original vocabulary $v$ and the set of special tokens $S$. These special tokens typically include markers for the start and end of sentences, unknown words, padding, etc., and are essential for certain SecurityBERT operations.

%------------------------------------------------
\subsubsection{Tokenizer Training}
%------------------------------------------------

\begin{equation}
T_d = \text{map}(v')
\end{equation}

Finally, the function $\text{map}(v')$ trains the tokenizer $T_d$. The trained tokenizer now maps future text inputs to the established vocabulary $v'$, effectively turning unstructured text into a form that the \texttt{SecurityBERT} can process. The trained tokenizer $T_d$ can take any text segment from the document $'d'$ into a series of tokens that \texttt{SecurityBERT} can understand.

These steps allow us to transform the raw text into a numerical representation that \texttt{SecurityBERT} can process effectively.

%----------------------------------------------------
\subsection{Fine tuning SecurityBERT}
%----------------------------------------------------

After the pre-training stage, we fine-tuned \texttt{SecurityBERT} for the cyber threat detection classification task. We added one linear layer followed by a Softmax activation function on top of the pre-trained \texttt{SecurityBERT} model, and the entire network is fine-tuned using our labeled data. This process enables \texttt{SecurityBERT} to adapt its learned contextual representations to the specific threat detection requirements, improving performance. 

%-------------------------------------------------
\subsubsection{Training setup}
%-------------------------------------------------

The training and fine-tuning were conducted on an Intel Xenon(R) 2.20 GHz CPU and an Nvidia A100 GPU with 40GB of RAM. The training on this specific hardware configuration was completed in 1 hour and 47 minutes. The following Section will extensively discuss a comprehensive performance evaluation of our novel \texttt{SecurityBERT} model.

%-------------------------------------------------
\subsection{Layers of the SecurityBERT model}
\begin{figure*}[htp]
    \centering
    \includegraphics[width=0.85\textwidth]{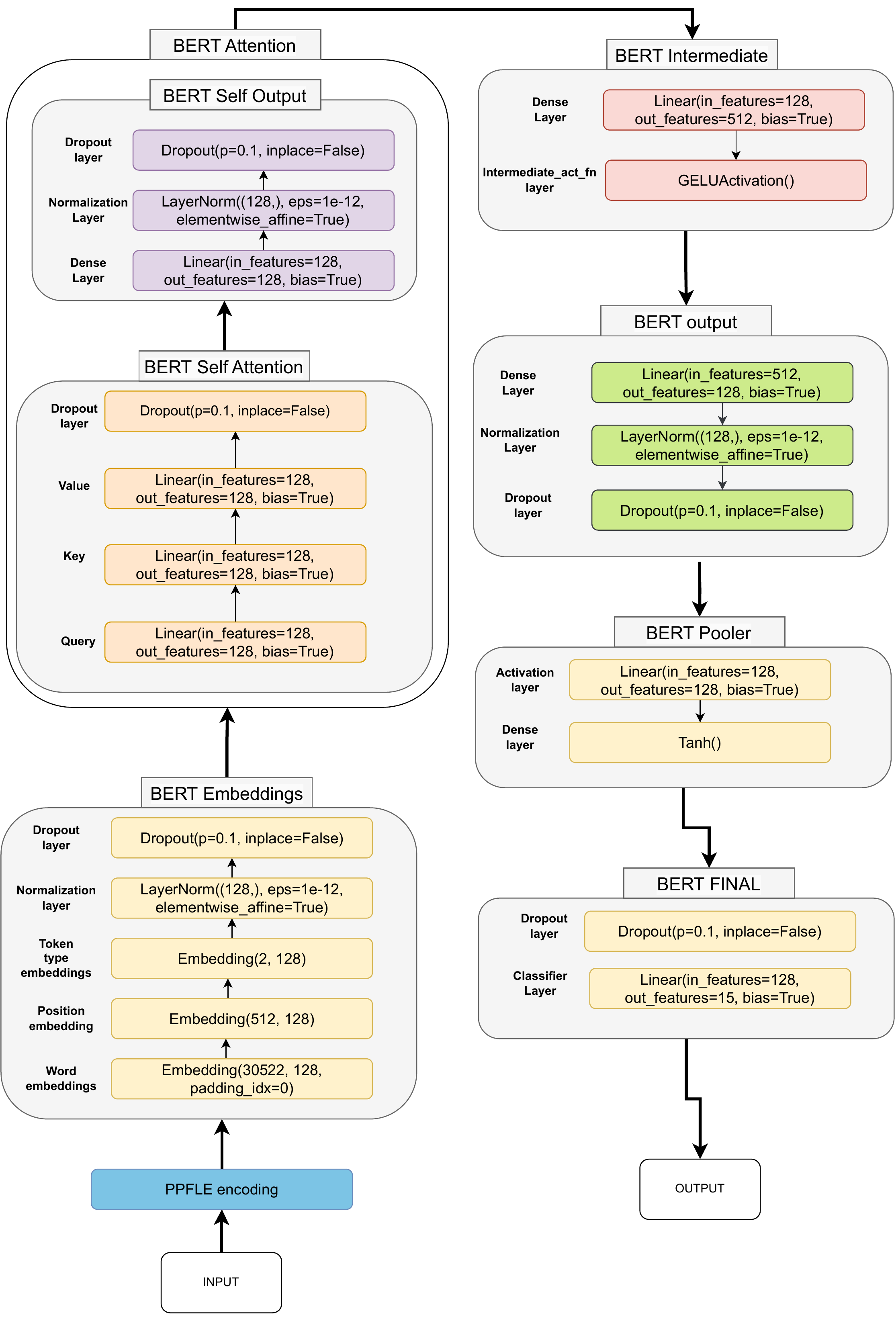}
  \caption{SecurityBERT architecture.}
    \label{fig:SecurityBERT}
\end{figure*}

%-------------------------------------------------
Throughout the research, the primary objective was to attain exceptional accuracy in data classification while ensuring the model’s size remained compact, with a focus on optimizing performance. After extensive experiments, the final model comprises 15 layers, specifically engineered to accurately comprehend PPFLE data while mitigating overfitting issues by incorporating suitable dropout layers. The comprehensive structure of the 15-layered \texttt{SecurityBERT} is illustrated in FIGURE~\ref{fig:SecurityBERT}. In the architectural design of \texttt{SecurityBERT}, we utilized just 4 Encoder Layers and modified the original parameters to suit our problem better. Additionally, we introduced a new layer in the final stage, comprising a Dropout layer, another new layer, and a Classifier Layer. TABLE~\ref{tab:bert_comparison} highlights the key parameter differences between the original two BERT models and \texttt{SecurityBERT}.

%-------------------------------------------------
\subsubsection{BERT Embeddings}
%-------------------------------------------------

The BERT Embeddings section starts with Word embeddings, succeeded by Position embeddings and then Token type embeddings. To stabilize the activations, there is a Layer Normalization with a size of 128, followed by a Dropout layer with a rate of 0.1.

%-------------------------------------------------
\subsubsection{BERT Self Attention}
%-------------------------------------------------
The BERT Self Attention comprises three primary linear transformations for the Key, Query, and Value. Each of these transformations has input and output features sized at 128. Another Dropout layer with a rate of 0.1 is included to prevent overfitting.

%-------------------------------------------------
\subsubsection{BERT Self Output}
%-------------------------------------------------
The BERT Self Output section features a Linear dense layer with an input and output feature size of 128. A Layer Normalization complements this, also sized at 128, and a Dropout layer with a rate of 0.1 for regularization.

%-------------------------------------------------
\subsubsection{BERT Intermediate}
%-------------------------------------------------
In the BERT Intermediate part, there's a dense layer with input features of 128 and output features expanded to 512. This section employs the GELU activation function.
%-------------------------------------------------
\subsubsection{BERT Output}
%-------------------------------------------------
In the BERT Output segment of the model, the final layer is a Linear dense layer that transforms the 512 features back to 128.
%-------------------------------------------------
\subsubsection{BERT Pooler+BERT Final}
%-------------------------------------------------
After a Tanh activation in the Bert Pooler, the output is streamlined through another Linear layer, further reducing the feature size to 15, representing the final output. This reduction is a crucial aspect of the model, preparing it for the 15 distinct classification tasks (14 attacks + 1 normal traffic).

%-----------------------------------------------------------
\subsection{Model Parameters}
%-----------------------------------------------------------

The precise parameter choices are among the most critical aspects of a BERT model. Incorrectly selected parameters can significantly influence the model's performance. The model uses a Byte-Pair Encoding Tokenizer, which provides a reliable and effective means of splitting input data into manageable tokens. The training data utilized by the model amounts to 661,767,168 tokens, with a limited vocabulary size of 5000. The minimum token frequency for \texttt{SecuriyBERT} is set at 2, while the model supports a maximum sequence length of 737 and a minimum sequence length of 619. The truncation settings limit the sequence length to a maximum 512, ensuring data consistency and model stability. Special tokens used by the model include \textless{}s\textgreater{}, \textless{}pad\textgreater{}, \textless{}/s\textgreater{}, \textless{}unk\textgreater{}, and \textless{}mask\textgreater{}. Regarding processing power, the model works with a batch size of 128 and a hidden size of 128. The model architecture comprises two hidden layers and utilizes four attention heads to process the input data. The intermediate size is set at 512, and the maximum position embeddings at 512, providing enough room for extensive and complex computations. The model can identify and respond to 14 different attacks, demonstrating its versatility and broad applicability. The \texttt{SecuriyBERT} is based on 11,174,415 parameters that are fine-tuned for optimal performance. Lastly, the model runs on an Nvidia A100 GPU, a powerful hardware accelerator that enables rapid data processing and real-time response capabilities.

TABLE~\ref{tab:tokenizer_config} summarizes the experimental parameter configuration, carefully designed to optimize performance and functionality. With the application of these parameters, our new \texttt{SecurityBERT} model exhibits the capability to identify fourteen distinct types of attacks with remarkable accuracy.

\begin{table}[ht!]
\centering
\caption{Configuration and parameres of SecurityBERT.}
\label{tab:tokenizer_config}
\begin{tabular}{|l|c|}
\hline
\rowcolor{lightgray}\textbf{Parameter}                   & \textbf{Value}                                                                                                                                               \\ \hline
Tokenizer type              & Byte-Pair Encoding Tokenizer                                                                                                                        \\ \hline
Training data               & 661,767,168 tokens                                                                                                                                  \\ \hline
Vocabulary size             & 5000                                                                                                                                                \\ \hline
Minimum token frequency     & 2                                                                                                                                                   \\ \hline
Maximum sequence length     & 737                                                                                                                                                 \\ \hline
Minimum sequence length     & 619                                                                                                                                                 \\ \hline
Truncation settings         & Max length=512                                                                                                                                      \\ \hline
Special tokens              & \textless{}s\textgreater{}, \textless{}pad\textgreater{},   \textless{}/s\textgreater{}, \textless{}unk\textgreater{},\textless{}mask\textgreater{} \\ \hline
Batch size                  & 128                                                                                                                                                 \\ \hline
Hidden size                 & 128                                                                                                                                                 \\ \hline
Number of hidden layers     & 2                                                                                                                                                   \\ \hline
Number of attention heads   & 4                                                                                                                                                   \\ \hline
Intermediate-size           & 512                                                                                                                                                 \\ \hline
Maximum position embeddings & 512                                                                                                                                                 \\ \hline
Number of attacks           & 14                                                                                                                                                  \\ \hline
Total number of parameters  & 11,174,415                                                                                                                                          \\ \hline
Hardware accelerator GPU    & Nvidia A100                                                                                                                                         \\ \hline
\end{tabular}
\end{table}

\section{Performance Evaluation of SecurityBERT}
\label{sec:sec4}
In this section, we evaluate the performance of the newly proposed \texttt{SecurityBERT}  model, through rigorous testing and comparative analysis. We show that the newly proposed model achieves a remarkable accuracy of $98.2$\%, which, to the best of our knowledge, stands as the highest accuracy ever attained using an ML algorithm detecting IoT attacks on realistic real-world network traffic.

%-------------------------------------------------
\subsection{Experimental Results}
\label{sec:experimental}
%-------------------------------------------------

To ensure appropriate comparisons with results from other models, we rely on standard measurements, namely Precision, Recall, F1-Score, and Support measurements. These metrics are crucial in comprehensively evaluating the model's performance and providing a meaningful assessment of its capabilities.

We partitioned the Edge-IIoTset dataset conventionally, allocating 80\% of the samples for training and reserving 20\% for evaluation. The model has not previously been exposed to the evaluation data, and we assess its effectiveness using those samples. TABLE~\ref{tab:datadis} presents the distribution of different types of cyber attack samples across training and evaluation data sets.

\begin{table}[ht]
\centering
\caption{Distribution of Data Across 14 Attack Types.} \label{tab:datadis}

%\scriptsize
\begin{tabular}{|l|r|r|r|}
\hline

\rowcolor{lightgray} \textbf{Attack Type} & \textbf{Nb. of Samples} & \textbf{Train. Data} & \textbf{Eval. Data}\\
\hline
Normal & 1,615,643 & 1,292,514 & 323,129 \\
DDoS\_UDP & 121,568 & 88,027 & 22,007 \\
DDoS\_ICMP & 116,436 & 93,149 & 23,287 \\
SQL\_injection & 51,203 & 40,962 & 10,241 \\
Password & 50,153 & 40,122 & 10,031 \\
Vulnerability\_scanner & 50,110 & 40,088 & 10,022 \\
DDoS\_TCP & 50,062 & 40,050 & 10,012 \\
DDoS\_HTTP & 49,911 & 39,929 & 9,982 \\
Uploading & 37,634 & 30,107 & 7,527 \\
Backdoor & 24,862 & 19,890 & 4,972 \\
Port\_Scanning & 22,564 & 18,051 & 4,513 \\
XSS & 15,915 & 12,732 & 3,183 \\
Ransomware & 10,925 & 8,740 & 2,185 \\
MITM & 1,214 & 320 & 80 \\
Fingerprinting & 1,001 & 801 & 200 \\\hline
Max Count & 2,219,201 & 1,765,482 & 441,371 \\
\hline
\end{tabular}
\end{table}

TABLE~\ref{tab:classi} presents the detailed classification report for the \texttt{SecuriyBERT} model on various network attack classes.

\begin{table}[ht!]
\centering
\caption{Classification Report of SecurityBERT.}\label{tab:classi}
\begin{tabular}{|l|c|c|c|c|}
\hline
\rowcolor{lightgray}\textbf{Class} & \textbf{Precision} & \textbf{Recall} & \textbf{F1-Score} & \textbf{Support} \\
\hline
Normal & 1.00 & 1.00 & 1.00 & 323,129 \\
DDoS\_UDP & 1.00 & 1.00 & 1.00 & 22,007 \\
DDoS\_ICMP & 1.00 & 1.00 & 1.00 & 23,287 \\
SQL\_injection & 0.85 & 0.83 & 0.84 & 10,241 \\
Password & 0.85 & 0.81 & 0.83 & 10,031 \\
DDoS\_TCP & 1.00 & 1.00 & 1.00 & 10,012 \\
DDoS\_HTTP & 0.89 & 0.99 & 0.94 & 9,982 \\
Vul\_scanner & 1.00 & 0.94 & 0.97 & 10,022 \\
Uploading & 0.79 & 0.86 & 0.83 & 7,527 \\
Backdoor & 0.82 & 0.94 & 0.87 & 4,972 \\
Port\_Scanning & 0.87 & 1.00 & 0.93 & 4,513 \\
XSS & 0.94 & 0.76 & 0.84 & 3,183 \\
Ransomware & 1.00 & 0.40 & 0.57 & 2,185 \\
Fingerprinting & 0.00 & 0.00 & 0.00 & 200 \\
MITM & 1.00 & 1.00 & 1.00 & 80 \\
\hline
Macro Avg & 0.87 & 0.84 & 0.84 & 441,371 \\ \hline
Weighted Avg & 0.98 & 0.98 & 0.98 & 441,371 \\ \hline
\textbf{Accuracy} & \multicolumn{4}{c|}{\textbf{0.982 (98.2\%)}} \\
\hline
\end{tabular}
\end{table}

\begin{figure*}[htp]
    \centering
    \includegraphics[width=0.8\textwidth]{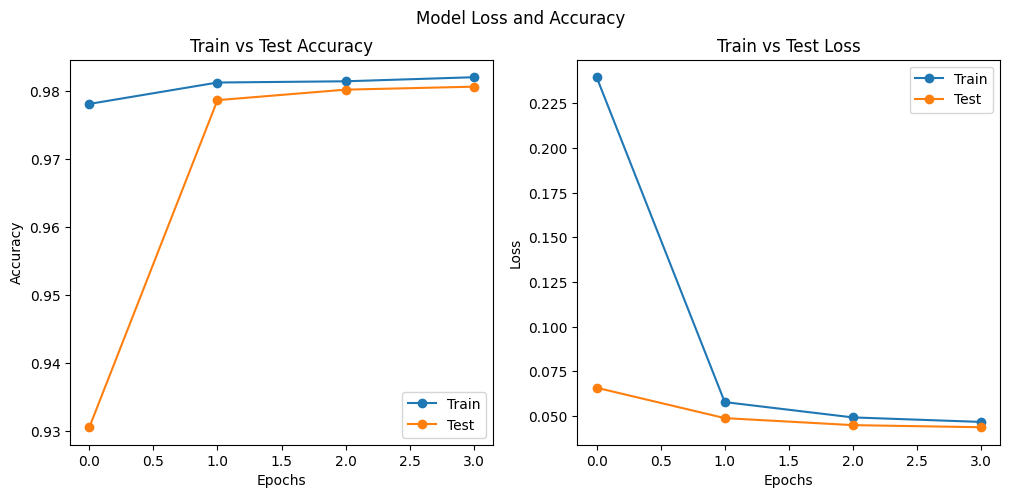}
  \caption{Accuracy and loss history of SecurityLLM training in 4 epochs.}
    \label{fig:trainin_loss}
\end{figure*}

FIGURE~\ref{fig:trainin_loss} shows the accuracy and loss history during the \texttt{SecuriyBERT} training changes over four epochs. 

%-------------------------------------------------
\subsubsection{ROC AUC Scores for Cyber Threat Classification}
%-------------------------------------------------

\begin{figure}[htp]
    \centering
 \includegraphics[width=.4\textwidth]{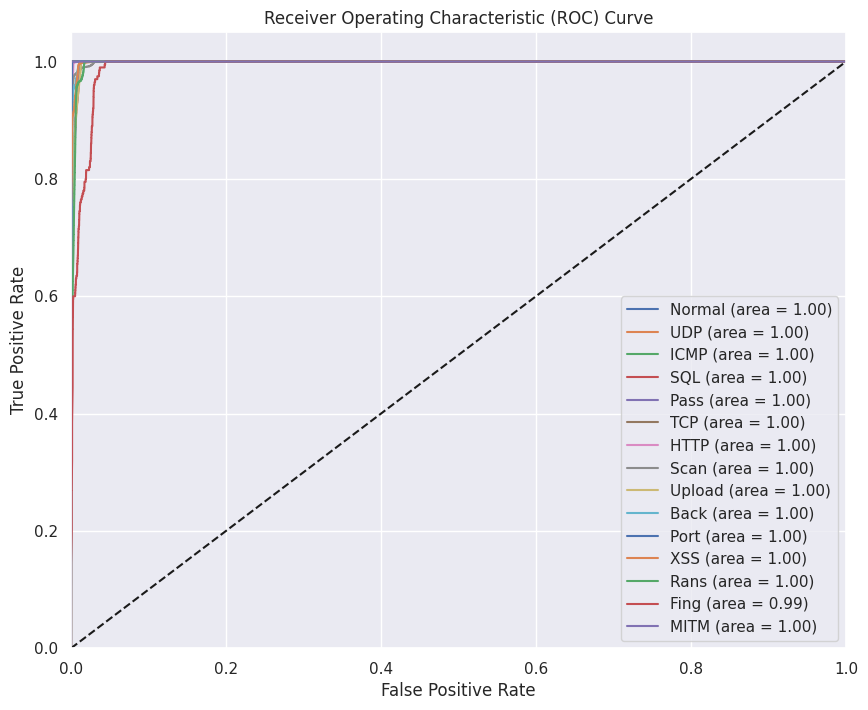}
  \caption{ROC AUC Scores for Cyber Threat Classification.}
    \label{fig:esd4}
\end{figure}

FIGURE~\ref{fig:esd4} presents various classes' Receiver Operating Characteristic Area Under the Curve (ROC AUC) scores. These scores indicate the \texttt{SecurityBERT} model's performance, with a value of 1.0 being perfect. Classes "Normal", "UDP", "TCP", and "MITM" demonstrate perfect classification with an AUC score of 1.0, which suggests that the model can flawlessly differentiate these classes from the others. The classes "ICMP", "SQL", "Pass", "HTTP", "Scan", "Upload", "Back", "Port", "XSS", and "Rans" all have AUC scores exceedingly close to 1.0, ranging from approximately 0.9976 to 0.999988. This implies a near-perfect classification for these classes, with very minor misclassifications. On the lower end of the performance spectrum, the class "Fing" has an AUC score of 0.991569, which, while still indicative of strong performance, means it has a slightly higher misclassification rate than the other classes. The \texttt{SecurityBERT} model generally exhibits stellar performance across all classes, with almost all of them achieving near-perfect or perfect classification.

%-------------------------------------------------
\subsubsection{Confusion Matrix}
%-------------------------------------------------

\begin{figure}[htp]
    \centering
 \includegraphics[width=.5\textwidth]{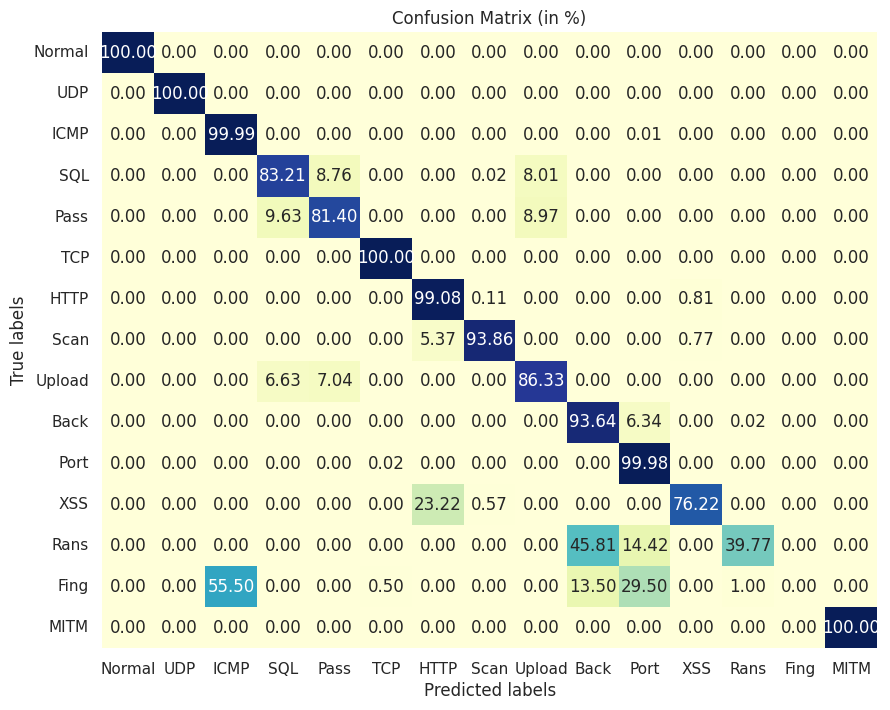}
  \caption{Confusion matrix of SecurityBERT classification.}
    \label{fig:conf_matrix}
\end{figure}

A visual depiction of the confusion matrix from the \texttt{SecuriyBERT} classification is presented in FIGURE~\ref{fig:conf_matrix}.

For the 'Normal' class and most types of DDoS attacks, including 'DDoS\_UDP', 'DDoS\_ICMP', and 'DDoS\_TCP', the model achieved perfect scores in terms of precision, recall, and F1-score, showing a high accurate classification performance on these types (c.f. TABLE~\ref{tab:classi}). It is noteworthy to mention the high support count for the 'Normal' class, which amounts to $323,129$ instances. The performance on 'SQL\_injection', 'Password', 'DDoS\_HTTP', 'Uploading', 'Backdoor', and 'Port\_Scanning' classes was relatively lower but still commendable, with F1-scores ranging from $0.83$ to $0.94$. Notably, 'DDoS\_HTTP' and 'Port\_Scanning' achieved a remarkably high recall of $0.99$ and $1.00$, respectively, indicating that the model could identify almost all instances of these attacks when they occur. 'Vul\_scanner' had a high precision and a slightly lower recall of $0.94$, resulting in an F1-score of $0.97$, showing good performance in identifying this type of attack. 'Ransomware' showed a high precision of $1.00$, but with a significantly lower recall of $0.40$, resulting in an F1-score of $0.57$, suggesting that while the model made correct predictions for the 'Ransomware' class, it missed a significant portion of actual instances. 

An examination of the confusion matrix reveals that, while the ransomware classification did experience misclassification in a substantial proportion of instances, most misclassifications occurred within the 'Backdoors' category. This category bears notable similarities with the ransomware category in real-life traffic data. Consequently, if our model misclassifies ransomware as a backdoor, it will not have a significant impact, maintaining satisfactory results in practical applications. The classes 'XSS' and 'MITM' showed good performance with F1-scores of $0.84$ and $1.00$, respectively, demonstrating that the model handled these classes well. Interestingly, the 'Fingerprinting' class had a precision, recall, and F1-score of 0, indicating a complete misclassification for these instances by the model. We again highlight that, much like the backdoor-ransomware misclassification scenario, a considerable proportion of the 'Fingerprint' misclassifications pertain to the 'ICMP' class. Misclassifying 'Fingerprint' as 'XSS,' for example, would ordinarily be a substantial issue in misclassification. However, in practical applications, these misclassifications bear no real consequence since the 'Fingerprint' and 'ICMP' classes closely resemble each other. 

The average recall and F1-score were all $0.84$ on the macro level. The weighted average was considerably higher at $0.98$ for all three metrics, suggesting a good performance overall. The slight difference between these two averages may be due to the imbalanced nature of the dataset, as classes with larger support have a greater influence on the weighted average. The overall accuracy of the model, measuring the proportion of correct predictions made out of all predictions, was $0.982$, showing a high degree of the predictive power of the \texttt{SecurityBERT} model in identifying different types of network attacks.  

\subsubsection{WeightWatcher - Empirical Spectral Density (ESD)}

\begin{figure}[htp]
    \centering
 \includegraphics[width=.4\textwidth]{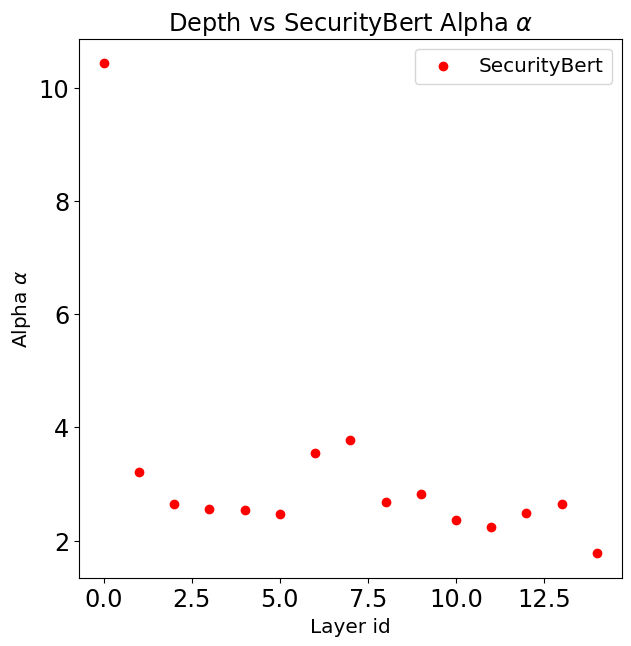}
  \caption{Power Law (PL) exponent ($\alpha$) values.}
    \label{fig:weight}
\end{figure}

\begin{figure}[htp]
    \centering
 \includegraphics[width=.4\textwidth]{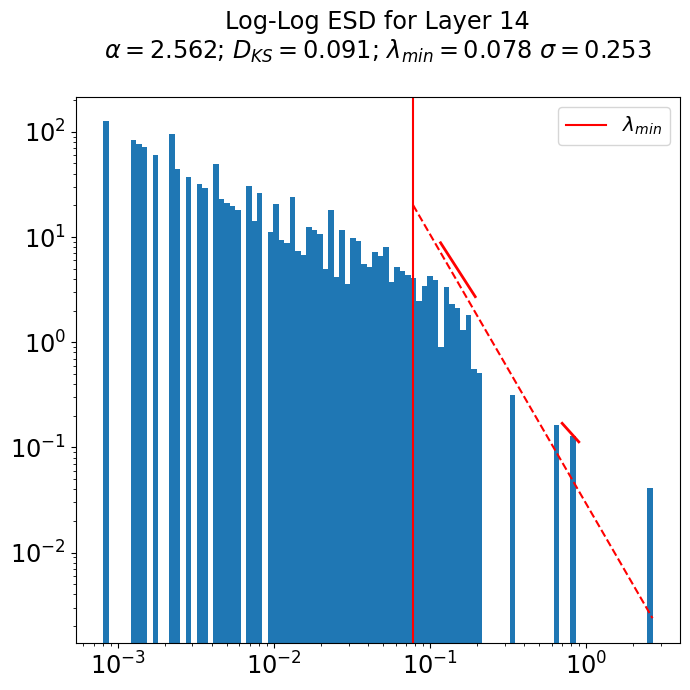}
  \caption{Empirical Spectral Density (ESD) for Layer 14.}
    \label{fig:esd}
\end{figure}

\begin{figure}[htp]
    \centering
 \includegraphics[width=.4\textwidth]{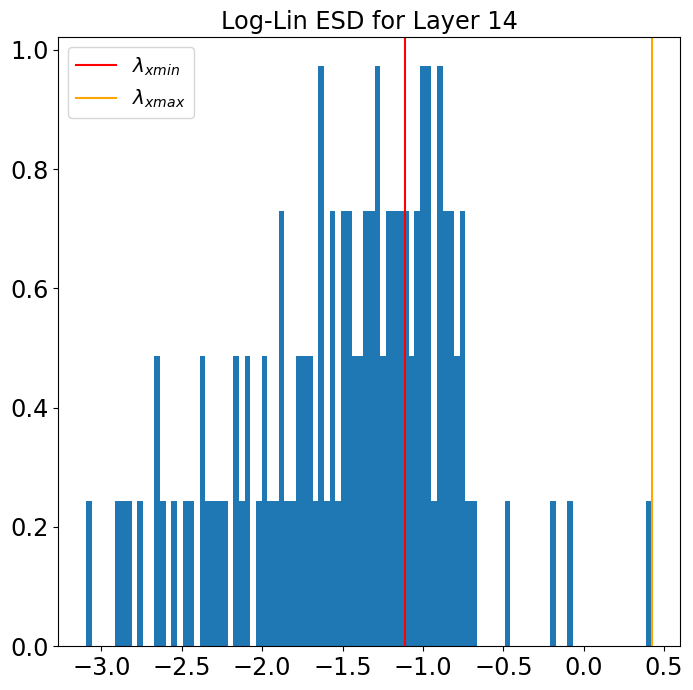}
  \caption{Log-Lin Empirical Spectral Density (ESD) for Layer 14.}
    \label{fig:esd3}
\end{figure}

WeightWatcher (WW) is an open-source diagnostic tool designed for examining Deep Neural Networks (DNNs) and can analyze various layers within a model. WeightWatcher can assist in identifying signs of overfitting and underfitting within particular layers of pre-trained or trained DNNs. We employed WW to optimize performance throughout our experiments, modifying the model's parameters to achieve optimal results.  FIGURE~\ref{fig:weight} presents the Power Law (PL) exponent ($\alpha$) values when plotted against layer IDs, revealing intriguing insights into the weight matrix properties of  \texttt{SecurityBERT}. FIGURE~\ref{fig:esd} presents the Empirical Spectral Density (ESD) for Layer 14. FIGURE~\ref{fig:esd3} presents the Log-Lin Empirical Spectral Density (ESD) for Layer 14.

Initial layers, especially the first, show a significantly high $\alpha$ value of around 10.43, suggesting a distinct weight initialization or early layer behavior. As we progress deeper into the network, the $\alpha$ values stabilize around 2 to 3, with many layers hovering close to the 2 mark. An $\alpha$ value near 2 indicates weight matrices possessing heavy-tailed properties, which, according to \cite{martin2021predicting}, smaller values ($\alpha$ $\approx 2$) are associated with models that generalize better.

According to the measurements presented, \texttt{SecurityBERT} can generalize effectively to new data that closely resembles the patterns observed during testing on the training dataset.

\begin{table*}[ht!]
\centering
\caption{Comparison of SecurityBERT with traditional ML and DL models.}\label{tab:comp}
\begin{tabular}{|c|c|c|c|c|}
\hline
\rowcolor{lightgray} \textbf{AI type} & \textbf{Authors} & \textbf{Year} & \textbf{AI Model} & \textbf{Accuracy} \\ \hline
\multirow{4}{*}{Traditional ML} & Ferrag et al.~\cite{ferrag2022edge} & 2022 & Decision Tree (DT)& 67.11\% \\ \cline{2-5}
& Ferrag et al.~\cite{ferrag2022edge}  & 2022 & Random Forest (RF) & 80.83\% \\ \cline{2-5}
& Ferrag et al.~\cite{ferrag2022edge}& 2022 & Support Vector Machines (SVM) & 77.61\% \\ \cline{2-5}
& Aouedi et al.~\cite{aouedi2023f}& 2023 & DT + RF / FL & 90.91\% \\ \cline{2-5}
&  Zhang et al.~\cite{zhang2023automatic} & 2023 & K-Nearest Neighbor (KNN)  & 93.78\% \\ \cline{2-5}
& Ferrag et al.~\cite{ferrag2022edge} & 2022 & K-Nearest Neighbor (KNN)  & 79.18\% \\ \hline
\multirow{7}{*}{Deep learning models} & Friha et al.~\cite{friha20232df} & 2023 & CNN / CL / No-DP  & 94.84\% \\ \cline{2-5}
 & Friha et al.~\cite{friha20232df} & 2023 & CNN / FL / No-DP  & 93.96\% \\ \cline{2-5}
  & Aljuhani et al.~\cite{aljuhani2023deep} & 2023 &  CSAE + ABiLSTM & 94.40\% \\ \cline{2-5}
& Friha et al.~\cite{friha2022felids} & 2022 & Recurrent Neural Network (RNN)& 94\% \\ \cline{2-5}
& Ding et al.~\cite{ding2023deepak} & 2023 & Long short-term memory (LSTM) & 94.96\% \\ \cline{2-5}
& Ferrag et al.~\cite{ferrag2022edge} & 2022 &  Deep Neural Network (DNN) & 94.67\% \\ \cline{2-5}
& Friha et al.~\cite{friha2022felids} & 2022 &  Deep Neural Network (DNN) & 93\% \\ \cline{2-5}
& E. M.d. Elias et al. \cite{de2022hybrid} & 2022 &  CNN-LSTM & 97.14\% \\ \cline{2-5}
  & Ferrag et al.~\cite{10181170} & 2023 & Transformer model w/o Tokenization and Embedding  & 94.55\% \\ \hline
\multirow{2}{*}{Large language model} & - & - & BERT without PPFLE & 51.3\% \\ \cline{2-5}
& This work & 2024 & SecurityBERT with PPFLE & \textbf{98.20\%} \\ \hline
\end{tabular}

CNN: Convolutional Neural Network, CL: Centralized Learning, FL: federated learning, DP: Differential Privacy, CSAE: Contractive Sparse AutoEncoder, ABiLSTM: Attention-based Bidirectional Long Short Term Memory.

\end{table*}

%-------------------------------------------------
\subsection{Performance Comparison}
%-------------------------------------------------

Numerous research studies have assessed the accuracy of detecting the 14 attacks in the Edge-IIoTset dataset. In this section, we have specifically analyzed research conducted by various authors.

The creators of the Edge-IIoTset dataset tested various traditional ML algorithms on it, including Decision Tree (DT), Random Forest (RF), Support Vector Machines (SVM), and K-Nearest Neighbor (KNN). Among these traditional methods, DT exhibited the lowest performance with an accuracy of only 67.11\%, while RF outperformed the others with an accuracy of 80.83\%. In addition to these traditional algorithms, a Deep Neural Network (DNN) test was conducted, which outperformed the others, boasting an accuracy of 94.67\%. The ultimate objective of this research is to develop a model capable of achieving nearly flawless real-time accuracy while maintaining a relatively compact model size suitable for deployment on IoT-embedded devices. This requirement explicitly rules out resource-intensive solutions like utilizing pre-trained LLMs, which, although capable of delivering high accuracy, are impractical for constrained devices regarding real-time packet analysis due to their significant resource demands. Following the initial dataset release, numerous authors tried to enhance accuracy using various model combinations and novel architectural designs. 
TABLE~\ref{tab:comp} presents the comparative accuracy of the proposed model, namely \texttt{SecurityBERT}, against the traditional ML models and Deep Learning (DL) models.

Friha et al.~\cite{friha20232df} explored the potential of CNNs to exceed the 95\% accuracy. They experimented with various setups, including Centralized Learning (CL) and Federated Learning (FL), both with and without Differential Privacy (DP). Using CL without DP, their best model attained an accuracy of 94.84\%. E. M.d. Elias et al. combined the CNN approach with Long-Short Term Memory (LSTM). This combination surpassed the 95\% threshold, reaching an impressive accuracy of 97.14\% on the Edge-IIoTset dataset, solely by extracting features from the transport and network layers.
A recent research paper by Ferrag et al.~\cite{Transformer95Ferrag} explored an innovative method. They introduced a simple GAN and Transformer-based architecture without any tokenization or embeddings. The model obtained a 94.55\% accuracy rate, and in this study, they raised the question of whether tokenization could pose challenges when applied to IoT datasets due to the unstructured nature of IoT network data, making it challenging to capture the semantics of closely resembling patterns like TCP scans or UDP scans.

The main objective of our paper was to create a novel model capable of exceeding an already exceptionally high level of accuracy.  To the best of our knowledge, we have reached a record-breaking accuracy of 98.2\% in classifying the 14 types of attacks, as showcased in TABLE~\ref{tab:classi}. This achievement represents the highest accuracy ever achieved in the multiclassification of these attack categories.

%-------------------------------------------------
\subsection{Real-life environment integration}
%-------------------------------------------------

\begin{figure}[htbp]
    \centering
    \includegraphics[width=0.25\textwidth]{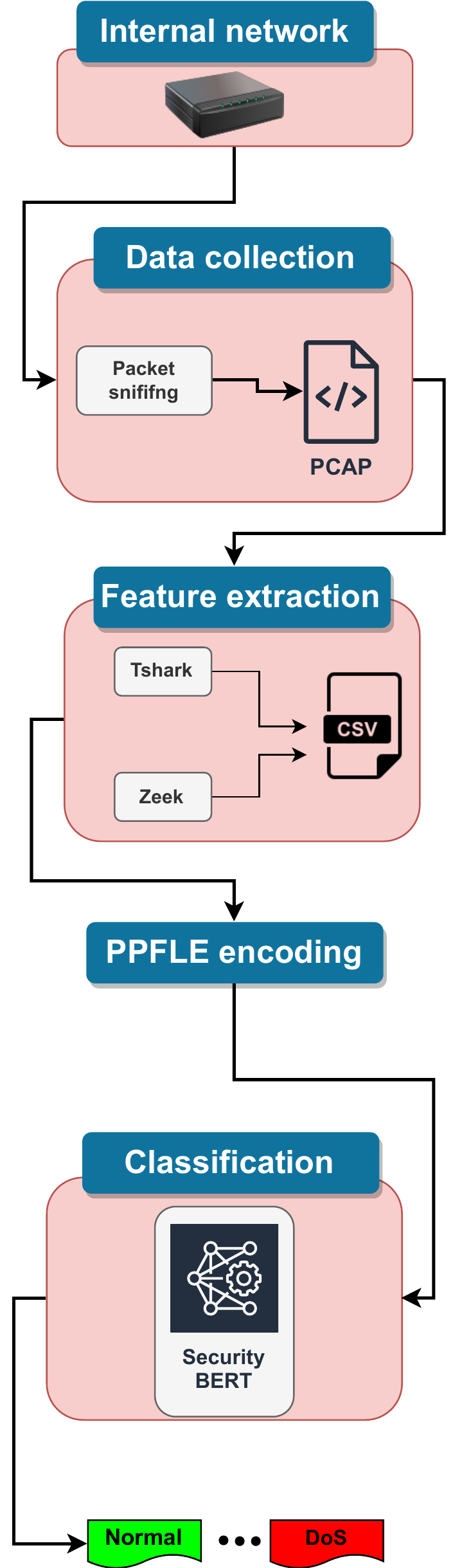}
  \caption{Real-life experimental setup using SecurityBERT.}
    \label{fig:realenvironment}
\end{figure}

In the original Edge-IIoTset dataset, feature extraction is derived from genuine PCAP files. This implies that replicating the same results in real-life scenarios becomes feasible if we possess real traffic data. We can substitute the PCAP files used in the Edge-IIoTset dataset~\cite{ferrag2022edge} with real-life internal network traffic, employing a suitable sniffing tool to generate the PCAP file. FIGURE~\ref{fig:realenvironment} provides a visual representation of the experimental setup where \texttt{SecurityBERT} is seamlessly integrated into a real-life network environment.
This system is designed to detect network incidents with remarkable accuracy, leveraging real-time network packet data.

%-----------------------------------------------------------
\subsubsection{Inference time}
%-----------------------------------------------------------

To implement the model on IoT devices, evaluating whether the inference time is sufficiently fast is essential. If the model is overly complex and exhibits slow inference times on an average CPU, its viability in real-world environments becomes questionable. TABLE~\ref{tab:hardware_performance} offers a detailed comparative analysis of computation times across different hardware platforms, specifically focusing on the inference task of the \texttt{SecuriyBERT} model.  The figures in the table denote the average inference time derived from $1000$ measurements.

\begin{table}[h!]
\centering
\begin{tabular}{|l|c|}
\hline
\rowcolor{lightgray}\textbf{Hardware} & \textbf{Average Inference Time (sec)} \\ \hline
A100 GPU & 0.0164 \\ \hline
T4 GPU & 0.0244 \\ \hline
V100 GPU & 0.0277 \\ \hline
TPU & 0.0327 \\ \hline
CPU* & 0.1582 \\ \hline
\end{tabular}\\
*Intel(R) Xeon(R) CPU @ 2.20GHz
\caption{Inference Time of SecuriyBERT Across Different Hardware Platforms}
\label{tab:hardware_performance}
\end{table}

\noindent The devices evaluated include three NVIDIA GPU models (A100, T4, and V100), Google's Tensor Processing Unit (TPU), and a general-purpose CPU. Each entry, denoted in seconds, reflects each hardware platform's time to perform the inference using \texttt{SecuriyBERT}. The A100 GPU is the most efficient in this context, completing the inference quickly. The pivotal metric here is the 0.15 sec CPU inference time, signifying that the model can be efficiently deployed on resource-limited devices for analyzing real-life traffic. Additionally, given its compact size of just 16.7 MB, the model is well-suited for deployment in embedded devices. 
\subsubsection{Reducing MTTR}
Integrating \texttt{SecurityBERT} into an embedded device and deploying it within an IoT network makes it possible to substantially enhance detection accuracy and leverage its high-speed performance to identify malicious activities within internal networks in real-time quickly. This, in turn, can lead to a notable reduction in Mean Time to Remediate (MTTR). 

Implementing AI in software security and incident handling is not a recent development in software security. For instance, companies like Rubrik\footnote{https://www.rubrik.com/products} and Microsoft have adopted generative AI models to optimize operations and enhance efficiency. For instance, if Rubrik's Security Cloud machine detects abnormal behavior, it automatically generates an incident in Microsoft's Sentinel. By employing this proactive approach, they can achieve faster response times and more effective management of potential security threats.

Similarly, \texttt{SecurityBERT} can be seamlessly integrated into existing real-world systems, thereby augmenting the overall accuracy and detection rate of these pre-existing systems.

%-----------------------------------------------------------
\section{Conclusion}
\label{sec:Conclusion}
%-----------------------------------------------------------

The innovative application of BERT architecture for cyber threat detection, embodied in \texttt{SecurityBERT}, demonstrated remarkable efficiency, contradicting initial assumptions regarding its incompatibility due to the reduced significance of syntactic structures. Experimental results underscored the superiority of this approach over conventional ML and DL models, including CNNs, deep learning networks, and recurrent neural networks. The \texttt{SecurityBERT} model, tested on a collected cybersecurity dataset, exhibited an outstanding capability to identify fourteen distinct types of attacks with an accuracy rate of $98.2$\%.

While this paper has made significant progress in advancing the use of LLMs in cybersecurity, future research directions can take several routes to enhance these promising findings further. One potential avenue involves fine-tuning and expanding the \texttt{SecurityBERT} model to enhance its performance across various attack types, incorporating adversarial attacks and more complex threats. In addition, due to the evolving nature of cyber threats, continuous updating and training of \texttt{SecurityBERT} model on the latest real-world datasets will be imperative to maintain its efficacy.

An exciting and promising avenue for future research is delving into methods to autonomously implement mitigations based on the classification of \texttt{SecurityBERT}. This advancement could lead to automated patch management, antivirus management, network reconfiguration, port management, and numerous other facets of cybersecurity management.

%=============================
%\section*{Acknowledgment}
%=============================

\bibliographystyle{unsrt}
\bibliography{ref}

\end{document}